\documentclass[journal]{IEEEtran}

\usepackage{cite}
\usepackage{amsmath,amssymb,amsfonts}
\usepackage{float}
\usepackage{graphicx}
\usepackage{textcomp}
\usepackage{xcolor}
\usepackage{url}
\usepackage{epstopdf}
\usepackage{algorithm,algorithmic}
    
\begin{document}

%
%
\title{The No-U-Turn Sampler as a Proposal Distribution in a Sequential Monte Carlo Sampler with a Near-Optimal L-Kernel}

\author{Lee Devlin, Paul Horridge, Peter L. Green, and Simon Maskell
\thanks{This work has been submitted to the IEEE for possible publication. Copyright may be transferred without notice, after which this version may no longer be accessible. Work funded by the Engineering and Physical Sciences Research Council, as part of the grant ‘Big Hypotheses: a Fully Parallelised Bayesian Inference Solution’ (EP/R018537/1)}
\thanks{Lee Devlin, Paul Horridge and Simon Maskell are with the Department of Electrical
Engineering and Electronics, and Peter Green is with the Department of Mechanical Engineering, University of Liverpool, Liverpool L69 3GJ, U.K.
(e-mail: Lee.Devlin@liverpool.ac.uk; Paul.Horridge@liverpool.ac.uk; plgreen@liverpool.ac.uk; S.Maskell@liverpool.ac.uk).}

}

\maketitle
 
%
%

\begin{abstract} 
Markov Chain Monte Carlo (MCMC) is a powerful method for drawing samples from non-standard probability distributions and is utilized across many fields and disciplines. Methods such as Metropolis-Adjusted Langevin (MALA) and Hamiltonian Monte Carlo (HMC), which use gradient information to explore the target distribution, are popular variants of MCMC. The Sequential Monte Carlo (SMC) sampler is an alternative sampling method which, unlike MCMC, can readily utilise parallel computing architectures and also has tuning parameters not available to MCMC. One such parameter is the L-kernel which can be used to minimise the variance of the estimates from an SMC sampler. In this letter, we show how the proposal used in the No-U-Turn Sampler (NUTS), an advanced variant of HMC, can be incorporated into an SMC sampler to improve the efficiency of the exploration of the target space. We also show how the SMC sampler can be optimized using both a near-optimal L-kernel and a Hamiltonian proposal.
\end{abstract}

\IEEEpeerreviewmaketitle

%
%
\section{Introduction}
Markov Chain Monte Carlo (MCMC) is a common tool used in Bayesian inference to draw samples from a probability distribution $\pi(\mathbf{x})$, where $\mathbf{x} \in \mathbb{R}^D$. Applications of MCMC span astronomy \cite{astronomy} to zoology \cite{Zoology}. MCMC involves moving to a state on the distribution $\mathbf{x}_k$ at iteration $k$ from a state $\mathbf{x}_{k-1}$ at the previous iteration, with some acceptance probability such that the Markov chain is ergodic (i.e., converges to a stationary distribution), and detailed balance is maintained, such that the stationary distribution of the Markov chain is equal to the target distribution. While many variations of MCMC exist, gradient based methods such as Metropolis-Adjusted Langevin (MALA)\cite{MALA} and Hamiltonian Monte Carlo (HMC)\cite{HMC} have rapidly grown in popularity due to their ability to efficiently explore continuous state spaces. HMC introduces a momentum vector $\mathbf{p} \in \mathbb{R}^D$ to explore states via the numerical integration of Hamiltonian dynamics and requires two parameters to be tuned to sample effectively. These are (i) the step-size taken by the numerical integrator, which must be fixed to satisfy detailed balance, and (ii) the number of steps taken between the start and end point of a trajectory. Tuning of the latter has been automated with the advent of the No-U-Turn Sampler (NUTS), first proposed in \cite{NUTS}, which calibrates the number of steps taken by stopping a trajectory once the path begins to turn back on itself. As a result of its applicability and efficient operation across a range of specific distributions, NUTS is used by popular probabilistic programming languages such as Stan \cite{Stan}, PyMC3\cite{pymc3} and NumPyro \cite{NumPyro}.\par 

Sequential Monte Carlo (SMC) samplers, first introduced in \cite{SMC}, provide a way of realising estimates based on a population of $N$ weighted hypotheses (often referred to as samples or particles) which evolve over $k$ iterations. In SMC, samples are mutated by means of a proposal distribution which moves the samples around the target space. While the proposal distribution used in the SMC literature is typically a Gaussian random-walk kernel, this is not a requirement. In this letter we show how the method NUTS uses to explore the target space can be utilised by an SMC sampler. We will refer to this new approach as SMC-NUTS.  Furthermore, we show how this can be leveraged with a near-optimal L-kernel, thus improving how the SMC sampler functions.

The rest of this letter is structured as follows. In Section \ref{SMC} we present how SMC samplers operate and in Section \ref{SMC-NUTS} we show how the proposal for NUTS can be used as the proposal distribution of an SMC sampler. Section \ref{Results} presents results in the context of two examples. Section \ref{conclusions} concludes the paper.

%
%
\section{Sequential Monte Carlo Samplers}
\label{SMC}
 In this work we consider an SMC sampler that does not target $\pi\left(\mathbf{x}\right)$ directly, but rather does so over $k$ iterations such that the joint distribution $\pi(\mathbf{x}_{1\ldots k})$ of all previous states is the target:
\begin{equation}
    \pi(\mathbf{x}_{1\ldots k}) = \pi(\mathbf{x}_{k}) \prod^{k}_{k'=2} L\left( \mathbf{x}_{k'-1} \vert \mathbf{x}_{k'} \right),
\end{equation}
\noindent where $L\left( \mathbf{x}_{k'-1} \vert \mathbf{x}_{k'} \right)$ is the L-kernel, also known as the 'backwards kernel', which is chosen such that:
\begin{equation}
    \int\pi(\mathbf{x}_{1\ldots k}) d\mathbf{x}_{1\ldots k-1}= \pi\left(\mathbf{x}_{k} \right).
\end{equation}

Using importance sampling we can attribute a weight to the $i^\mathrm{th}$ hypothesis at iteration $k$, $w_{k}^{i}$, which is updated from the previous iteration's weight $w_{k-1}^{i}$ via: 
\begin{equation}
\label{weights}
    w_{k}^{i}=w_{k-1}^{i} \cdot \frac{\pi(\mathbf{x}_{k}^{i})}{\pi(\mathbf{x}_{k-1}^{i})} \cdot \frac{L(\mathbf{x}_{k-1}^{i}|\mathbf{x}_{k}^{i})}{q(\mathbf{x}_{k}^{i}|\mathbf{x}_{k-1}^{i})},
\end{equation}
\noindent where $\pi(\mathbf{x}_{k}^{i})$ and $\pi(\mathbf{x}_{k-1}^{i})$ are the target evaluated at the $i^\mathrm{th}$ sample's new and previous state, respectively, and $q(\mathbf{x}_{k}^{i}|\mathbf{x}_{k-1}^{i})$ is the forwards proposal distribution. The forwards proposal distribution and L-kernel are probability distributions associated with transformations to the state $\mathbf{\mathbf{x}}_{k}^{i}$ from the state $\mathbf{\mathbf{x}}_{k-1}^{i}$, and vice versa. $L(\mathbf{x}_{k-1}^{i}|\mathbf{x}_{k}^{i})$ can be any valid probability distribution function and it is sometimes convenient to set  $L(\mathbf{x}_{k-1}^{i}|\mathbf{x}_{k}^{i}) = q(\mathbf{x}_{k-1}^{i}|\mathbf{x}_{k}^{i}) $. We note, however, that such an approach may be sub-optimal and that the appropriate selection of $L(\mathbf{x}_{k-1}^{i}|\mathbf{x}_{k}^{i})$ allows the variance of the estimates to be minimised  \cite{PG}. 

When degeneracy occurs, where a small subset of samples have relatively high importance weights, a new set of samples are selected from the current set with probability proportional to the normalised weight, $\widetilde{w}$, in a process called resampling. This involves selecting elements, with replacement, from $[\mathbf{x}_k^1 \ldots \mathbf{x}_k^N]$ with probability $[\widetilde{w}^1_k\ldots \widetilde{w}^N_k]$ into a new vector $\mathbf{x}_k^{\mathrm{new}}$ which then overwrites the old samples, i.e. $\mathbf{x}_k =\mathbf{x}_k^{\mathrm{new}}$, before the weights of the new samples are all set to $1/N$. Resampling is typically set to occur when the effective number of samples falls below some threshold value, usually half the total number of samples. \par

Values of interest, e.g. expected values with respect to the target distribution, can be realised from the normalised sample weights, and furthermore, it is possible to use previous estimates to improve the current estimate at iteration $k$ using recycling schemes \cite{recycling}. 

%
%

\section{The No-U-Turn sampler as a proposal distribution of an SMC sampler} \label{SMC-NUTS}

When used in MCMC, NUTS generates samples from a proposal of the form $q(\mathbf{x}_{k}, \mathbf{p}_{k} | \mathbf{x}_{k-1}, \mathbf{p}_{k-1})$. In an SMC sampler, to calculate (\ref{weights}), we wish to consider a proposal of the form $q(\mathbf{x}_{k}^{i}|\mathbf{x}_{k-1}^{i})$. We can address this disparity by considering the numerical integration of the Hamiltonian dynamics to be a non-linear function which transforms  samples to a new position by means of the momentum. 

HMC and NUTS use a numerical method called Leapfrog to simulate Hamiltonian dynamics and explore the target space. Leapfrog has several useful qualities. Firstly it is symplectic, i.e. it preserves the geometric structure of the phase space $\{\mathbf{x},\mathbf{p}\}$, and therefore generates states with high acceptance probability for sufficiently small step-sizes. Secondly, it is both reversible and time symmetric such that detailed balance is maintained. The Leapfrog method over one step of step-size $h$ is as follows:
\begin{align} 
    &\mathbf{p}_{k-\frac{1}{2}} = \mathbf{p}_{k-1}-\frac{h}{2}\left. \frac{\partial U}{\partial \mathbf{x}}\right|_{\mathbf{x}_{k-1}}   \\
    &\mathbf{x}_{k}= \mathbf{x}_{k-1} + h\boldsymbol{M}^{-1} \mathbf{p}_{k-\frac{1}{2}}   \\
    &\mathbf{p}_{k}= \mathbf{p}_{k-\frac{1}{2}} -\frac{h}{2}\left. \frac{\partial U}{\partial \mathbf{x}} \right|_{\mathbf{x}_{k}} 
\end{align}
\noindent where $U$ is a potential energy function and is related to the target distribution by $U(\mathbf{x})=-\log(\pi(\mathbf{x}))$, and $\boldsymbol{M} \in \mathbb{R}^{D \times D}$ is a diagonal mass matrix. 

%
%

\subsection{Non-linear transform of the proposal distribution}
\label{sec:proposals}

We wish to evaluate the probability that a random variable $\mathbf{X}_{k-1}$
transforms to a random variable $\mathbf{X}_{k}$ using a Hamiltonian based
proposal, and vice-versa. We write this as $q(\mathbf{X}_{k} = \mathbf{x}_{k}  |\mathbf{X}_{k-1} = \textbf{x}_{k-1})$ for the forwards kernel and $L(\mathbf{X}_{k-1} = \mathbf{x}_{k-1}  |\mathbf{X}_{k} = \textbf{x}_{k})$ for the L-kernel. 

First we derive an expression for the forwards kernel. 
We generalise Leapfrog to a single function $f_{LF}$(.) which transforms a state $\mathbf{x}_{k-1}$ to $\mathbf{x}_{k}$, i.e. $\mathbf{x}_{k}=f_{LF}(\mathbf{x}_{k-1},\mathbf{p}_{k-1})$. We can rewrite the forward kernel as:
\begin{align}
&q(\mathbf{X}_{k}=\mathbf{x}_{k}|\mathbf{X}_{k-1}=\mathbf{x}_{k-1}) = \nonumber\label{transform}\\ &q(\mathbf{X}_{k}=f_{LF}(\mathbf{x}_{k-1},\mathbf{p}_{k-1})|\mathbf{X}_{k-1}=\mathbf{x}_{k-1}).
\end{align}

In a Hamiltonian proposal the momentum term is the stochastic term which changes the value of $\mathbf{x}_{k}$. We can therefore write (\ref{transform}) in terms of a random momentum variable $\mathbf{P}$ by using a change of variables as follows:
\begin{align}
&q(\mathbf{X}_{k}=f_{LF}(\mathbf{x}_{k-1},\mathbf{p}_{k-1})|\mathbf{X}_{k-1}=\mathbf{x}_{k-1}) =  \nonumber\\ &q(\mathbf{P}_{k-1}=\mathbf{p}_{k-1}|\mathbf{X}_{k-1}=\mathbf{x}_{k-1}) \left|\frac{df_{LF}(\mathbf{x}_{k-1},\mathbf{p}_{k-1})}{d\mathbf{p}_{k-1}}\right|^{-1}.
\end{align}
 
 The initial velocity is typically sampled from a normal distribution $\mathbf{p}_k \sim \mathcal{N}\left(0,\boldsymbol{M}\right)$, we therefore find that:
\begin{align}
&q(\mathbf{X}_{k}=\mathbf{x}_{k}|\mathbf{X}_{k-1}=\mathbf{x}_{k-1}) = \nonumber \\
&\mathcal{N}(\mathbf{p}_{k-1};0,\boldsymbol{M})\left|\frac{df_{LF}(\mathbf{x}_{k-1},\mathbf{p}_{k-1})}{d\mathbf{p}_{k-1}}\right|^{-1}.
\label{Kkernel}
\end{align}

To evaluate the L-kernel we utilise the fact that Leapfrog is a reversible integration method, i.e. if we start at a state $\{\mathbf{x}_{k-1},\mathbf{p}_{k-1}\}$ and then transform this to $\{\mathbf{x}_{k},\mathbf{p}_{k}\}$ then by applying $\mathbf{-p}_{k}$ it follows: $\mathbf{x}_{k-1}=f_{LF}(\mathbf{x}_{k},-\mathbf{p}_{k})$. Following the same steps to arrive at (\ref{Kkernel}), except this time we start at $\mathbf{x}_{k}$ and with a a velocity -$\mathbf{p}_{k}$ we arrive at:
\begin{align}
&L(\mathbf{X}_{k-1}=f_{LF}(\mathbf{x}_{k},\mathbf{-p}_{k})|\mathbf{X}_{k}=\mathbf{x}_{k}) = \nonumber \\  &L(\mathbf{P}_{k}=-\mathbf{p}_{k}|\mathbf{X}_{k}=\mathbf{x}_{k}) \left|\frac{df_{LF}(\mathbf{x}_{k},\mathbf{-p}_{k})}{d\mathbf{p}_{k}}\right|^{-1}.
\label{Lkernel}
\end{align}

 For each sample in an SMC iteration to calculate (\ref{weights}) we need to calculate the ratio of (\ref{Lkernel}) and (\ref{Kkernel}). As we will now explain, it is fairly straightforward to show that the determinant terms cancel when Leapfrog is used. Writing the updated state in terms of the initial state and momentum we find that:  
\begin{equation}
\label{eq:det}
\left| \frac{\partial f_{LF}(\mathbf{x}_{k-1},\mathbf{p}_{k-1})}{\partial \mathbf{p}_{k-1}} \right|
    = h^D \prod_{i=1 }^{D} \boldsymbol{M}^{-1}_{ii}. 
\end{equation}
As Leapfrog is a reversible method, if the the momentum is reversed and the step-size is equal to that used in the forwards case we similarly find:
\begin{equation}
\left| \frac{\partial f_{LF}(\mathbf{x}_{k},-\mathbf{p}_{k})}{\partial \mathbf{p}_{k}} \right|
    = h^D \prod_{i=1 }^{D} \boldsymbol{M}^{-1}_{ii}. 
\end{equation}
\noindent such that the determinants will cancel when calculating (\ref{weights}).
For a reversible method, like Leapfrog, the integrator will evaluate at all the states that were previously evaluated when going forwards. As such, when using the proposal from NUTS, the determinants cancel when calculating the second fraction term in (\ref{weights}).

%
%

\subsection{A near-optimal L-kernel for an SMC sampler with a NUTS proposal}

We are free to choose the distribution of the L-kernel. One approach is to assume the reverse of the forward proposal, i.e. we can assume that $\mathbf{p}_{k}$ is sampled from the same distribution as the proposal for the initial momentum:
\begin{equation}
\label{symmetricL}
    L(\mathbf{P}_{k}=-\mathbf{p}_{k}|\mathbf{X}_{k}=\mathbf{x}_{k}) = \mathcal{N}(-\mathbf{p}_{k};\mathbf{0}, \boldsymbol{M}). 
\end{equation}

This approach, however, is sub-optimal. The optimal L-kernel $L^{\mathrm{Opt}}$ is that which minimises the variance of the sample estimates. It can be shown that $L^{Opt} \propto   q(\mathbf{x}_{k}|\mathbf{x}_{k-1}) \eta (\mathbf{x}_{k-1})$  where $\eta(\mathbf{x}_{k-1})$ is the distribution of samples at iteration $k-1$ \cite{SMC} and, as seen in Section \ref{sec:proposals},  for the Hamiltonian case it follows that $L^{\mathrm{Opt}} \propto q(-\mathbf{p}_{k}|\mathbf{x}_{k-1}) \eta (\mathbf{x}_{k-1})$. Closed-form expressions for the optimal L-kernel are often intractable. To approximate the optimal L-kernel we follow the approach given in \cite{PG}, and for the Hamiltonian case take the distribution of the negative new momentum and new samples,  $\eta(\mathbf{P}_{k}=-\mathbf{p}_{k})q(\mathbf{X}_k=\mathbf{x}_k|\mathbf{P}_k=\mathbf{p}_k)$, and form a Gaussian approximation such that: 
\begin{align}
    &\eta(\mathbf{P}_{k}=-\mathbf{p}_{k})q(\mathbf{X}_k=\mathbf{x}_k|\mathbf{P}_k=\mathbf{p}_k) \approx \nonumber \\ &\mathcal{N}\left( \begin{bmatrix} -\mathbf{p}_{k}\\ \mathbf{x}_{k} \end{bmatrix} ; \begin{bmatrix} \boldsymbol{\mu}_{-\mathbf{p}_{k}}\\ \boldsymbol{\mu}_{\mathbf{x}_{k}} \end{bmatrix}, \begin{bmatrix} \boldsymbol{\Sigma}_{-\mathbf{p}_{k},-\mathbf{p}_{k}} & \boldsymbol{\Sigma}_{-\mathbf{p}_{k},\mathbf{x}_{k}} \\ \boldsymbol{\Sigma}_{\mathbf{x}_{k},-\mathbf{p}_{k}} & \boldsymbol{\Sigma}_{\mathbf{x}_{k},\mathbf{x}_{k}} \end{bmatrix} \right),
    \label{eq::joint}    
\end{align}
\noindent where $\boldsymbol{\mu} \in \mathbb{R}^D$ are mean vectors, and $\boldsymbol{\Sigma} \in \mathbb{R}^{D \times D}$ are block covariance matrices.

We then use the properties of Gaussians to define a near-optimal L-kernel:
\begin{equation}
  L^{\mathrm{Opt}}(\mathbf{P}_{k}=-\mathbf{p}_{k}|\mathbf{X}_{k}=\mathbf{x}_{k}) \approx \mathcal{N}(-\mathbf{p}_{k} ; \boldsymbol{\mu}_{-\mathbf{p}_{k}|\mathbf{x}_{k}}, \boldsymbol{\Sigma}_{-\mathbf{p}_{k}|\mathbf{x}_{k}}),
    \label{eq}
\end{equation}
\noindent where: 
\begin{equation}
\label{eq2}
    \boldsymbol{\mu}_{-\mathbf{p}_{k}|\mathbf{x}_{k}}= \boldsymbol{\mu}_{-\mathbf{p}_{k}} + \boldsymbol{\Sigma}_{-\mathbf{p}_{k},\mathbf{x}_{k}} \boldsymbol{\Sigma}_{\mathbf{x}_{k},\mathbf{x}_{k}}^{-1}(\mathbf{x}_{k} - \boldsymbol{\mu}_{\mathbf{x}_{k}} )
\end{equation}
\noindent and
\begin{equation}
\label{eq3}
    \boldsymbol{\Sigma}_{-\mathbf{p}_{k}|\mathbf{x}_{k}}=\boldsymbol{\Sigma}_{-\mathbf{p}_{k},-\mathbf{p}_{k}} - \boldsymbol{\Sigma}_{-\mathbf{p}_{k},\mathbf{x}_{k}} \boldsymbol{\Sigma}_{\mathbf{x}_{k},\mathbf{x}_{k}}^{-1}\boldsymbol{\Sigma}_{\mathbf{x}_{k},-\mathbf{p}_{k}}.
\end{equation}

 Returning to the incremental weights (\ref{weights}), updates are found via:
\begin{equation}
    w^{i}_k=w^{i}_{k-1} \cdot \frac{\pi(\mathbf{x}_k)}{\pi(\mathbf{x}_{k-1})} \cdot \frac{L(\mathbf{P}_{k}=-\mathbf{p}_{k}|\mathbf{X}_{k}=\mathbf{x}_{k})}{q(\mathbf{P}_{k-1}=\mathbf{p}_{k-1}|\mathbf{X}_{k-1}=\mathbf{x}_{k-1})},
    \label{OptWUpdate}
\end{equation}

\noindent where $\mathbf{x}_{k-1}$ and $\mathbf{p}_{k-1}$ are the initial position and momentum, and $\mathbf{x}_k$ and $\mathbf{-p}_{k}$ are the position and negative momentum after a NUTS iteration. The right numerator can be evaluated from either (\ref{symmetricL}) or (\ref{eq}) and the right denominator can be evaluated from the initial momentum distribution. Algorithm \ref{pseudo} shows how the near-optimal L-kernel is used within SMC-NUTS for $N$ samples over a total of $T$ iterations. Algorithm 3 in \cite{NUTS} can be used to generate new samples for the NUTS step. For the sub-optimal symmetric L-kernel, steps 10 and 11 are replaced by (\ref{symmetricL}). For the resampling step, several methods may be employed, see \cite{AV} and references therein for a discussion on resampling in the context of particle filters (which may be applied here).

 \begin{algorithm}
 \caption{SMC-NUTS with a near-optimal L-kernel for $T$ iterations and $N$ samples.}
 \label{pseudo}
 \begin{algorithmic}[1]
  \FOR{i=1\ldots N}
      \STATE Sample $\mathbf{x}^{i}_{1}$ from  $q(\mathbf{x}^{i})$ 
      \STATE Set initial weights to $w^{i}_{1} = \frac{\pi(\mathbf{x}^{i}_{1})}{q(\mathbf{x}^{i}_{1})}$ 
  \ENDFOR
  \FOR {$k = 2$ to $T$}
  \FOR{i=1\ldots N}
    \STATE Sample an initial momentum vector $\mathbf{p}^{i}\sim \mathcal{N}(\mathbf{0},\mathbf{M})$
    \STATE $(\mathbf{x}^{i}_{k},\mathbf{p}^{i}_{k})= NUTS(\mathbf{x}^{i}_{k-1},\mathbf{p}^{i})$
    \ENDFOR
    \STATE Calculate parameters of (\ref{eq::joint})
    \STATE Calculate (\ref{eq}) using (\ref{eq2}) and (\ref{eq3}).
    \FOR{i=1\ldots N}
        \STATE Update sample weights $w_k^i$ using (\ref{OptWUpdate})
    \ENDFOR
    \FOR{i=1\ldots N}
        \STATE Calculate normalised weights: $\widetilde{w}^{i}_k =\frac{w^{i}_k}{\boldsymbol{\Sigma}_{j=1}^{j=N} w^j_k}$
    \ENDFOR
    \STATE Calculate effective number of samples: \\ $N_{eff} = \frac{1}{\Sigma_{j=1}^{j=N}{\widetilde{w}^{j2}_k}}$ 
    \IF {$N_{eff} < N/2$}
    
                \STATE Resample $ [ \mathbf{x}_k^1 \ldots \mathbf{x}_k^{N}]$ with probability $[\widetilde{w}^1_k\ldots \widetilde{w}^N_k]$ \;
                \STATE Reset all weights to $\frac{1}{N}$ \;
    \ENDIF
    \ENDFOR
 \end{algorithmic}
 \end{algorithm}

%
%
\section{Results}
\label{Results}
We now demonstrate SMC-NUTS in two example cases. In both examples the mass matrix is set equal to an identity matrix.

\subsection{Penalised Regression with Count Data}
We first compare our method against a state-of-the-art SMC sampler with a random walk proposal which was used to estimate parameters of a penalised regression model with count data in \cite{Nguyen}. \par

This problem makes use of Lasso regression \cite{LASSO} whereby a penalty constraint $\gamma \sum^D_{j=1} \left| \mathbf{\beta}_j\right|$ is placed on the size of the regression coefficients $\mathbf{\beta}\in \mathbb{R}^D$. We follow \cite{Nguyen} (with associated details from \cite{NguyenPhD}) by using the exponential power distribution bridge framework for our regularizing prior:
\begin{equation}
    f(\mathbf{\beta}; \gamma,z)=\prod^D_{j=1}\frac{z}{2\gamma\Gamma(1/z)}\exp \left( - \left| \frac{\mathbf{\beta}_j}{\gamma}\right|^z\right),
\end{equation}
\noindent where $z \in (0,2)$.
Our aim is to estimate coefficients used to generate count data. The likelihood is a Poisson distribution $\mathbf{y}_i \sim p(\mathbf{y}_i|\mathbf{\mu}_i)$ for the $i^{\mathrm{th}}$ observation, where:
\begin{equation}
 \mathbf{\mu}_{i}=\exp \left(\mathbf{\beta}_{0} + \sum^D_{j=1}\mathbf{\beta}_{j}\Phi^{j}_{i}\left(\mathbf{x}_{i,j}\right) \right).   
\end{equation}

To aid comparison with \cite{Nguyen}, we likewise generate 100 observations with a 12-Dimensional $\beta$ vector where $\beta_{0}=1$, $\beta_{2}=1.5$, $\beta_{4}=-2$, $\beta_{6}=1$, $\beta_{7}=-2$, $\beta_{9}=1.2$ and all other values are set to zero. The basis function $\Phi$ is a Gaussian kernel $\Phi=\exp \left( -\frac{(\mathbf{x}_i-\mathbf{c}_j)^2}{2\mathbf{r}^2_j}\right)$ with 11 equispaced centres $\mathbf{c}_j$ and all $\mathbf{r}_j$ values set to 0.5. In our case, we run the SMC-NUTS based proposal but, unlike the method in \cite{Nguyen}, choose not to run the sampler with any form of tempering  or step adaption. However, we do use the \cite{Nguyen}'s recycling scheme. Furthermore, we use $ \mathcal{N}(-\mathbf{p}_{k}; \mathbf{0}, \boldsymbol{I})$ as our L-kernel (i.e. we investigate the benefits of the SMC-NUTS proposal without the use on an approximately optimal L-kernel).\par

Table \ref{Tab:Penalised_Regression_results} shows the mean-squared-error (MSE) using SMC-NUTS compared to the random walk approach \cite{NguyenPhD}, where results were available, for different numbers of samples and iterations for $z=0.5$. The MSE for SMC-NUTS is 13-18 times smaller than for SMC with a random walk proposal for the same number of samples and iterations. 

\begin{table}
\centering
\caption{Average mean squared error for estimated model parameters for an SMC sampler using a NUTS and results from \cite{NguyenPhD} using a random walk proposal.}
\label{Tab:Penalised_Regression_results}
\begin{tabular}{l|ccc|ccc}
                     & \multicolumn{3}{c|}{T=100} & \multicolumn{3}{c}{T=200} \\
                     & N=25     & N=50        & N=200       & N=25        & N=50         & N=200       \\ \hline \hline
Random Walk           &  -       & 5.92        & 5.17        &  -          & 5.49         & 4.9         \\
NUTS                 &  0.51    & 0.471       & 0.420       & 0.324       & 0.306        & 0.267     
\end{tabular}
\end{table}

It should be noted that the NUTS proposal is more computationally expensive than a random walk proposal. This is demonstrated in Table \ref{Tab:Penalised_Regression_time} where we compare the average time it takes for SMC with NUTS to complete compared to SMC with a basic random walk proposal, i.e. with no cooling strategy, covariance adaption, or parallel implementation. While the NUTS variant takes longer to complete for an equal number of samples and iterations, with fewer samples and fewer iterations, the runtime is more competitive and still achieves substantially better accuracy. This is exemplified in the case of 25 samples for 100 iterations with a NUTS proposal which was quicker than 200 samples for 200 iterations with a random walk and achieves almost 10 times smaller mean-squared-error. The random walk algorithm used to generate results for Table \ref{Tab:Penalised_Regression_results} is more complex that the algorithm used to generate Table \ref{Tab:Penalised_Regression_time} and, therefore, our runtime for the random walk is likely to be an underestimation.

\begin{table}
\centering
\caption{Average runtime in seconds for an SMC sampler with a NUTS proposal and a random walk proposal simpler than that in \cite{Nguyen}}
\label{Tab:Penalised_Regression_time}
\begin{tabular}{l|ccc|ccc}
                     & \multicolumn{3}{c|}{T=100} & \multicolumn{3}{c}{T=200} \\
                     & N=25         & N=50        & N=200       & N=25        & N=50        & N=200        \\ \hline \hline
Random Walk          & 0.235        & 0.898       & 0.461       & 0.234       & 0.461       & 1.85         \\
NUTS                 & 1.50         & 3.08        & 13.40       & 3.05        & 6.41        & 24.0      
\end{tabular}
\end{table}

\subsection{Multivariate Student-t Distribution}
\begin{figure}
    \centering
    \includegraphics[width=0.48\textwidth]{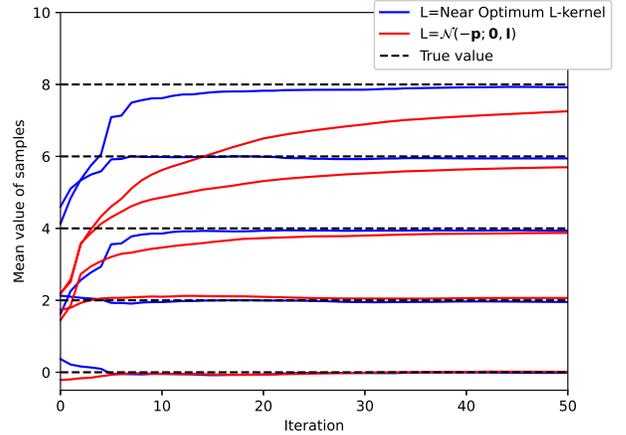}
    \caption{Mean of samples in each dimension using a near optimal L-kernel and a single Gaussian L-kernel targeting a 5-dimensional Student-t distribution. Black dashed lines show the true value.}
    \label{fig:student_t}
\end{figure}

In this example we aim to show how a near optimal L-kernel can improve the performance of the SMC-NUTS sampler.
We compare the simple symmetric L-kernel we used in the previous example with a near optimal L-kernel and purposefully initialise the samples away from the target's probability mass. For our target distribution we choose a Student's-t distribution, i.e: 
\begin{equation}
   \pi(x)= \frac{\Gamma \left(\frac{\nu+1}{2} \right)} {\sqrt{\nu\pi}\,\Gamma \left(\frac{\nu}{2} \right)} \left(1+\frac{x^2}{\nu} \right)^{-\frac{\nu+1}{2}}, 
\end{equation}
where $\nu$ is the number of degrees of freedom (set to 5 in this example). We set the mean vector of the distribution to $\mu=\left[ 0, 2, 4, 6, 8 \right]$ and run the SMC-NUTS for both L-kernel strategies for 200 samples and 50 iterations. The estimation of the mean vector $\mu$ is shown in Fig \ref{fig:student_t}. 

Using the near-optimal L-kernel it is clear that, through the use of an approximately optimal L-kernel, the SMC sampler is able to find the true values more rapidly than when using the symmetric L-kernel. This shows that it is possible to run the sampler for fewer iterations and achieve better results.

%
%

\section{Conclusions}
\label{conclusions}
We have shown how the proposal from the No-U-Turn Sampler (NUTS) can be used as a proposal in an SMC sampler. This allows efficient exploration of a wide variety of distributions, using gradient information whilst at the same time having the benefits SMC samplers offer in terms of being readily parallelisable. We have also shown that SMC-NUTS can benefit from the use of a near optimal L-kernel. An SMC sampler utilising NUTS as a proposal distribution is observed to give better results compared to a Gaussian proposal. It has also been demonstrated that it is possible to realise estimates that converge to true values in fewer iterations when this approach is taken. We recommend further work to extend the near optimal L-kernel to be applicable in a wider class of high-dimensional settings than have been considered to date (e.g. n \cite{PG}) such that SMC can be applied in contexts where NUTS is likely to deliver significant benefit.

\section*{Acknowledgment}
We wish to thank Dr Alexander Phillips for supplying feedback on our implementation of the penalized count data example.

\end{document}